\documentstyle[12pt,epsfig]{article}

\newcommand{\be}{\begin{equation}}
\newcommand{\ee}{\end{equation}}
\newcommand{\Dlt}{\Delta}
\newcommand{\dlt}{\delta}

\newcommand{\bt}{\beta}
\newcommand{\vp}{\varphi}
\newcommand{\ep}{\varepsilon}
\newcommand{\al}{\alpha}
\newcommand{\ra}{\rightarrow}

\begin{document}

\begin{center}

{\Large {\bf Self-similar factor approximants for evolution equations and
boundary-value problems} \\ [5mm]

E.P. Yukalova$^1$, V.I. Yukalov$^2$, and S. Gluzman$^3$} \\ [3mm]

{\it

$^1$Department of Computational Physics, Laboratory of Information
Technologies,\\
Joint Institute for Nuclear Research, Dubna 141980, Russia \\ [3mm]

$^2$Bogolubov Laboratory of Theoretical Physics, \\
Joint Institute for Nuclear Research, Dubna 141980, Russia \\ [3mm]

$^3$Generation 5 Mathematical Technologies Inc., \\
Corporate Headquaters, 515 Consumers Road, Toronto, Ontario M2J
4Z2, Canada}

\end{center}

\vskip 2cm

\begin{abstract}

The method of self-similar factor approximants is shown to be very
convenient for solving different evolution equations and boundary-value
problems typical of physical applications. The method is general and simple,
being a straightforward two-step procedure. First, the solution to an
equation is represented as an asymptotic series in powers of a variable.
Second, the series are summed by means of the self-similar factor
approximants. The obtained expressions provide highly accurate approximate
solutions to the considered equations. In some cases, it is even possible
to reconstruct {\it exact} solutions for the whole region of variables,
starting from asymptotic series for small variables. This can become
possible even when the solution is a transcendental function. The method
is shown to be more simple and accurate than different variants of
perturbation theory with respect to small parameters, being applicable
even when these parameters are large. The generality and accuracy of the 
method are illustrated by a number of evolution equations as well as 
boundary value problems.

\end{abstract}

\vskip 1cm

{\bf PACS numbers}: 02.30.Hq, 02.30.Mv, 02.60.Lj, 03.75.Lm, 05.45.Yv

\vskip 1cm

{\bf Keywords}: Evolution equations; Boundary-value problems; Solitons;
Bose-Einstein condensate; Vortices

\newpage

\section{Introduction}

Differential equations appear in numerous problems of physics,
applied mathematics [1--3] and many other branches of natural as
well as social sciences (see, e.g., [4--7]). In the majority of
cases, these equations are nonlinear and cannot be solved exactly,
allowing for exact solutions only in a few exceptional instances.
Then, in order to obtain approximate analytical solutions, one
resorts to perturbation theory in powers of some parameters assumed
to be small [1--3]. The resulting expressions are usually rather
cumbersome and are difficult to analyze. Also, they form asymptotic
series that are useful only for very small expansion parameters.

The validity of perturbative series can be extended to the finite
values of parameters by reorganizing them with the help of the
optimized perturbation theory [8]. The basic idea of this theory
is to include in the initial approximation a set of auxiliary
parameters which are transformed, at each step of perturbation
theory, into control functions governing the series convergence
[8]. The optimized perturbation theory has been successfully
applied to a great variety of problems, providing rather accurate
approximations (see review article [9] and references therein).
It has also been applied to solving differential equations [10-17].
However, the weak point of this approach to finding the solutions
of differential equations is that the optimization procedure results
in very complicated equations for control functions, which are to
be solved numerically. It is practically always much easier to solve
the given differential equations numerically than to deal with the
cumbersome optimization equations for control functions, anyway
requiring numerical solution. This is why this approach, though
being very useful for many other problems [9], has not found wide
practical use for solving differential equations.

Another method of constructing approximate solutions is based on the
self-similar approximation theory [18--26]. Then the solutions to
differential equations can be represented in the form of self-similar
exponential approximants or self-similar root approximants [27--34].
These approximants represent well those functions whose behavior at
large variables is known to be either exponential or power-law,
respectively.

In the present paper, we advocate a novel approach to constructing
approximate solutions of differential equations. This approach is
based on the use of {\it self-similar factor approximants} [35--37].
The mathematical derivation of the latter also rests on the
self-similar approximation theory [18--26], but the structure of
these approximants is rather different from the exponential and
root approximants [27--34]. The structure of the self-similar factor
approximants reminds that of thermodynamic characteristics near
critical points [38,39]. This is why it has been natural to apply,
first, these approximants to the description of critical phenomena
[35--37,40]. It was shown that these approximants allow us a
straightforward and simple determination of critical points and
critical indices, agreeing well with the results of the most
complicated numerical techniques, whose description can be found
in articles [41--46] and books [47--49].

The self-similar factor approximants make it possible to
define an effective sum of divergent series. Initially, these
approximants were introduced [35--37] for summing the partial
series of even orders, while the summation of odd-order series was
not defined. Recently, the method was completed by defining the
factor approximants of odd orders [40,50]. Now we have in hands
a general and uniquely prescribed procedure for constructing the
self-similar factor approximants of arbitrary orders. We show
below that this procedure can be employed for finding very
accurate approximate solutions to differential equations. In
some cases, the method gives exact solutions, if these exist.
The approach is very general, being applicable to linear as well
as to nonlinear equations, and to initial-value as well as to the
boundary-value problems. The principal difference of the method
from other perturbation theories [1--3] is that we, first,
represent the sought solutions as an asymptotic series in powers
of the equation variable, but not in powers of a parameter. And,
second, we extrapolate the given series to the whole range of the
variable by means of the self-similar factor approximants. The
advantage of the method is its extreme simplicity combined with
the high accuracy of the obtained solutions.

\section{Self-similar factor approximants}

Here we describe the general procedure of constructing self-similar
factor approximants as solutions to differential equations. We keep
in mind ordinary differential equations, though the procedure can
be generalized to partial differential equations. It would be
unreasonable to plunge from the very beginning to complicated
matters, but the principal idea should, first, be illustrated by
not too complex problems. In line with this, we consider the case,
where the solution is a real function of a real variable.
Thus, we assume that we are interested in finding a solution to a
differential equation, which is a real function $f(x)$ of a real
variable $x$. First, we are looking for the solution in the region
of asymptotically small $x\ra 0$, where it can be represented as an
asymptotic series
\be
\label{1}
f(x) \simeq f_0(x) \; \sum_n a_n\; x^n \qquad (x \ra 0) \; ,
\ee
where $f_0(x)$ is a given function. It is worth emphasizing that
the expansion variable $x$ is not compulsory the given independent
variable, but can be a function of it, which means that one can
always accomplish the change of variables and look for the expansion
in powers of the new variable. For instance, if the initial variable
entering an equation is $t$, we are not obliged to look for an
expansion necessarily in powers of $t$, but we can make a change
of the variable, introducing $x=x(t)$ and to study an expansion in
powers of $x(t)$. In particular, $x(t)$ could be $t^\al$. If $\al$
is not an integer, we get the Puiseux expansion. In general, one
can consider asymptotic expansions of the form $\sum_n a_nx_n$, with
functions $x_n$ from the given comparison scale [51].

Suppose, series (1) is limited by a finite sum of $k$-th order
\be
\label{2}
f_k(x) = f_0(x) \; \sum_{n=0}^k a_n x^n \; ,
\ee
with $k=0,1,2,\ldots$. Without the loss of generality, we may
set $a_0=1$, since if $a_0$ would not be 1, we could include it
in $f_0(x)$.

The self-similar factor approximants of even orders $k=2p$, with
$p=1,2,\ldots$, are defined [35--37] as
\be
\label{3}
f_{2p}^*(x) = f_0(x) \prod_{i=1}^p \left ( 1 +
A_i x \right )^{n_i} \; .
\ee
The parameters $A_i$ and $n_i$ are obtained from expanding Eq. (3)
in powers of $x$ and comparing the resulting expansion with sum (2)
by the accuracy-through-order procedure. The corresponding equations
for $a_i$ and $n_i$ can be written as
\be
\label{4}
\sum_{i=1}^p n_i A_i^n = B_n \qquad (n=1,2,\ldots,2p) \; ,
\ee
where
\be
\label{5}
B_n = \frac{(-1)^{n-1}}{(n-1)!}\; \lim_{x\ra 0} \;
\frac{d^n}{dx^n} \; \ln \; \sum_{m=0}^k \; a_m x^m \; .
\ee
Equations (4), whose number is $2p$, define all $2p$ unknowns
$A_i$ and $n_i$. The factor approximants (3) extrapolate expansion
(2), valid for small $x\ra 0$, to the region of finite $x$.

For odd orders $k=2p+1$, with $p=0,1,2,\ldots$, the factor
approximants are defined [40,50] in the form
\be
\label{6}
f_{2p+1}^*(x) = f_0(x) \; \prod_{i=1}^{p+1} \left ( 1 + A_i x
\right )^{n_i} \; ,
\ee
in which $A_1=1$ and all other $2p+1$ unknown parameters are given
by $2p+1$ equations
\be
\label{7}
\sum_{i=1}^{p+1} n_i A_i^n = B_n \qquad (n=1,2,\ldots,2p+1) \; ,
\ee
with $B_n$ from Eq. (5).

The index $k$, labelling the factor approximants $f_k^*(x)$, is
determined by the order $k$ of $f_k(x)$ in sum (2), which is used
for calculating the parameters $A_i$ and $n_i$. It may happen that
there exist additional conditions imposed on the behavior of the
sought function $f(x)$. Then the structure of $f_k^*(x)$ should
include additional factors, whose parameters are chosen so that to
satisfy the imposed conditions. In the case of differential equations,
such additional restrictions are given by initial and boundary
conditions.

For example, the Cauchy initial value problems are supplemented
by the set of initial conditions for the sought function and its
derivatives
\be
\label{8}
f(0) = f_0 \; , \qquad \left [
\frac{d^m f(x)}{dx^m} \right ]_{x=0} = f_0^{(m)} \; ,
\ee
where $m=1,2,\ldots$. Therefore, the factor approximants are
constructed so that to satisfy the same initial conditions
\be
\label{9}
f_k^*(0) =  f_0 \; , \qquad
\left [ \frac{d^m f_k^*(x)}{dx^m} \right ]_{x=0} = f_0^{(m)}
\ee
for all orders of $k$.

In the case of a boundary value problem, for a function $f(x)$ on an
interval of $x\in[x_1,x_2]$, one has the boundary conditions
\be
\label{10}
f(x_1) = f_1 \; , \qquad f(x_2) = f_2 \; .
\ee
Respectively, the factor approximants of each order have to satisfy
the same boundary conditions
\be
\label{11}
f_k^*(x_1) = f_1 \; , \qquad f_k^*(x_2) = f_2 \; .
\ee

When a boundary value problem is formulated for an infinite interval,
the related boundary conditions are given in the form of the asymptotic
behavior of the function $f(x)$. For instance, let the asymptotic form
of $f(x)$ at large $x$ be prescribed as
\be
\label{12}
f(x) \simeq Bx^\bt \qquad (x\ra\infty) \; ,
\ee
where $B\neq 0$ and $\bt$ is any real quantity, including zero.
Hence, the factor approximants $f_k^*(x)$ of any order must possess
the same asymptotic form
\be
\label{13}
f_k^*(x) \simeq B x^\bt \qquad (x\ra\infty) \; .
\ee
Assume that the zero-order factor in Eqs. (1) and (2) behaves at large
$x$ as
\be
\label{14}
f_0(x) \simeq c x^\al \qquad (x\ra\infty) \; ,
\ee
with $c\neq 0$ and any real $\al$. Then, in order that the factor
approximants (3) and (6) would satisfy the asymptotic boundary
condition (13), it should be that
\be
\label{15}
c \prod_i A_i^{n_i} =  B \; , \qquad \al + \sum_i n_i = \bt \; .
\ee

Depending on whether initial-value or boundary-value problems
are considered, we shall always require that the corresponding
conditions be valid {\it exactly}. This distinguishes our approach
from perturbation theories with respect to some parameters, when
boundary conditions are usually satisfied only approximately [1--3].

\section{Initial value problems}

Considering the Cauchy initial value problems, we shall concentrate
our attention on the so-called singular problems, where a parameter
enters in front of the highest derivative [1--3]. When one tries to
apply perturbation theory with respect to the given parameter, one
confronts nonuniformly valid expansions [1--3]. However in our
approach, we start with an expansion not in powers of this parameter,
but in powers of the equation variable. Then, using the extrapolation
by self-similar factor approximants, we obtain accurate solutions for
any value of the parameter. Moreover, in some cases, the factor
approximants restore {\it exact} solutions. This fact is so unusual
that we illustrate it in detail below.

\vskip 5mm

{\bf A. Linear singular problem}

\vskip 3mm

Let us consider the singular initial value problem
\be
\label{16}
\ep \; \frac{d^2 y}{dt^2} + 2\; \frac{dy}{dt} +
\frac{y}{\ep} = 0 \; ,
\ee
with the initial conditions
\be
\label{17}
y(0) = 1 \; , \qquad \left ( \frac{dy}{dt} \right )_{t=0} = -\;
\frac{1}{\ep} \; ,
\ee
defining $y=y(t)$ for $t\geq 0$. And let us pretend that we do not
know its exact solution.

Following our approach, we, first, look for the behavior of the
solution $y(t)$ at asymptotically small $t\ra 0$, which is
\be
\label{18}
y(t) \simeq 1 \; - \; \frac{t}{\ep} \; + \; \frac{t^2}{2\ep^2} \; +
\ldots \qquad (t\ra 0) \; .
\ee
For the factor approximant of second order, we get
\be
\label{19}
y_2^*(t) = \lim_{A\ra 0} ( 1 + At)^{-1/A\ep} = e^{-t/\ep} \; ,
\ee
which is the exact solution of Eq. (16). Similarly, for the factor
approximants of higher orders after $k=3,4,5$ and so on, we find
\be
\label{20}
y^*_k(t) = e^{-t/\ep} \qquad (k\geq 3) \; ,
\ee
that is, the {\it exact} solution of Eq. (16) under the initial
conditions (17) and any real $\ep$.

It is easy to notice that function (20) cannot in principle be
expanded in terms of $\ep$. The functions of this type are called
instanton solutions [47,52]. Note also that the exponential (20)
is a transcendental function.

\vskip 5mm

{\bf B. Nonlinear singular problem}

\vskip 3mm

The model problem of carrier transfer [2,3] is given by the
nonlinear equation
\be
\label{21}
( \ep y + t)\; \frac{dy}{dt} + y - 1 = 0
\ee
for $y=y(t)$ and $t\geq 0$, with the initial condition
\be
\label{22}
y(0) = 2 \; .
\ee
We again pretend that we are not aware of the solution to Eq.
(21). If we try to invoke perturbation theory with respect to
the parameter $\ep$, we come to a rather nontrivial singular
problem. The method of strained coordinates or the method of
asymptotic matching could be used, but with a necessity of
dealing with very lengthy and cumbersome calculations [2,3].
We show here that this problem is easily solvable by our method.

As is explained in Sec. 2, it is always possible to make a change
of variables, which seems to be convenient. If we introduce the
function $z(x)$, such that
\be
\label{23}
z = x + y \; , \qquad x = \frac{t}{\ep} \; ,
\ee
then Eq. (21) simplifies to
\be
\label{24}
z \; \frac{dz}{dx} \; - \; x \; - \; 1 = 0 \; ,
\ee
with the initial condition
\be
\label{25}
z(0) = 2 \; .
\ee
The asymptotic behavior of $z(x)$ at small $x$ is
\be
\label{26}
z(x) \simeq 2 + \frac{1}{2}\; x + \frac{3}{16}\; x^2 + \ldots
\qquad (x\ra 0) \; .
\ee

Constructing the factor approximants, we find that, starting from
the fourth order, the approximants $z_k^*(x)$ provide the exact
solutions to Eq. (24), with the initial condition (25). Thus, in
fourth order, we have
\be
\label{27}
z_4^*(x) = 2 ( 1 + A_1 x)^{n_1} ( 1 + A_2 x)^{n_2} \; .
\ee
From the accuracy-through-order procedure, with the given initial
condition (35), we get
\be
\label{28}
A_1 = \frac{1}{4}\left ( 1 - i\sqrt{3}\right ) \; , \qquad
A_2 = A_1^* \; , \qquad n_1 = n_2 = \frac{1}{2}\; .
\ee
Thence Eq. (27) reduces to
\be
\label{29}
z_4^*(x) = \sqrt{4+2x+x^2} \; ,
\ee
which is the exact solution of Eq. (24). Acting in the same way
for higher-order approximants, and using relations (23), we come
to the factor approximants
\be
\label{30}
y_k^*(t) = \sqrt{ 4 + \frac{2t}{\ep} + \frac{t^2}{\ep^2} } \;
- \; \frac{t}{\ep} \qquad (k\geq 4) \; ,
\ee
giving the {\it exact} solution of Eq. (21) for orders $k\geq 4$,
any real $\ep$, and all $t\geq 0$.

\vskip 5mm

{\bf C. Singular logistic equation}

\vskip 3mm

The logistic equation and its variants are widely employed in the
studies of population dynamics [53]. For the population $p(t)$,
being a function of time $t$, with the growth rate $1/\ep$, the
logistic equation writes as
\be
\label{31}
\ep \; \frac{dp}{dt} = p (1-p) \; .
\ee
It is given for $t\geq 0$, with the initial condition
\be
\label{32}
p(0) = p_0 \; .
\ee
If again we pretend not to know the solution of Eq. (31) and try
to resort to perturbation theory in powers of the large growth rate,
that is, small $\ep$, we confront a singular value problem.

Wishing to apply our approach, it is reasonable, as has been
discussed above, to choose the most convenient variables. The hint
on what variables would be the most convenient comes from analyzing
the dynamics in the vicinity of the stationary points. Equation (31)
possesses one stable fixed point $p(\infty)=1$. The motion near this
point is given by the asymptotic law
\be
\label{33}
p(t) \simeq 1 + c e^{-t/\ep} \qquad (t\ra \infty) \; .
\ee
The latter suggests to choose as the variable
\be
\label{34}
x \equiv e^{-t/\ep} \; .
\ee
Considering the population $p(t)$ as a function of this variable
(34), for the function
\be
\label{35}
y = p(t(x)) = y(x) \; ,
\ee
we have the equation
\be
\label{36}
\frac{dy}{dx} = y ( y - 1) x \; ,
\ee
with the initial condition
\be
\label{37}
y(1) = p_0 \; .
\ee

The small $x$-expansion of $y(x)$ is
\be
\label{38}
y(x) \simeq \sum_n a_n \; x^n \qquad (x \ra 0 ) \; ,
\ee
for which Eq. (36) yields $a_n=a_1^n$. The lowest-order factor
approximant, for which the initial condition can be satisfied,
is the third-order approximant resulting in the expression
\be
\label{39}
y_3^*(x) = ( 1 + A_1 x)^{n_1} \; ,
\ee
where
$$
A_1 = \frac{1}{p_0} \; - \; 1 \; , \qquad n_1 = -1 \; .
$$
The same expression (39) follows for other approximants of higher
orders $k\geq 3$. Returning to the population function $p(t)$ by
means of relations (34) and (35), we find
\be
\label{40}
p_k^*(t) = \frac{p_0}{p_0-(p_0-1)e^{-t/\ep}} \qquad (k\geq 3) \; .
\ee
This is the {\it exact} solution of the logistic equation (31),
with the initial condition (32).

\section{Boundary value problems}

In the previous section, we have shown that the method of
self-similar factor approximants allows us to reconstruct exact
solutions of some differential equations representing initial
value problems. This highly nontrivial fact occurs as well for
some nonlinear boundary value problems.

\vskip 5mm

{\bf A. Kink soliton equation}

\vskip 3mm

Nonlinear equations possessing soliton solutions are met in various
problems of physics and applied mathematics. Let us consider, for
example, the nonlinear Schr\"odinger equation describing the so-called
$\vp^4$-model with particle mass $1/\ep$. The equation reads as
\be
\label{41}
\frac{\ep}{2} \; \frac{d^2\vp}{dx^2} + \vp - \vp^3 = 0 \; .
\ee
Assume that the function $\vp=\vp(x)$ satisfies the boundary conditions
\be
\label{42}
\vp(-\infty) = -1 \; , \qquad \vp(\infty) = 1 \; .
\ee
From Eq. (41) and the boundary conditions (42) it follows that
$\vp(x)$ is an antisymmetric function, such that
\be
\label{43}
\vp(-x) = - \vp(x) \; ,
\ee
and, therefore,
\be
\label{44}
\vp(0) = 0 \; .
\ee

To choose a convenient expansion variable, we again study the form
of the function $\vp(x)$ in the vicinity of the stable stationary
points $\vp(\mp\infty)=\mp 1$, where we have
$$
\vp(x) \simeq - 1 + a \exp\left ( \frac{2}{\sqrt{\ep}} \; x
\right ) \qquad ( x \ra -\infty ) \; ,
$$
\be
\label{45}
\vp(x) \simeq 1 - a \exp\left ( -\; \frac{2}{\sqrt{\ep}} \; x
\right ) \qquad ( x \ra \infty ) \; ,
\ee
with $a$ being a real parameter. This immediately suggests to choose
as a variable
\be
\label{46}
z \equiv \exp \left ( \frac{2}{\sqrt{\ep}}\; x \right ) \; .
\ee
In terms of the latter variable, Eqs. (45) become
$$
\vp(x) \simeq - 1 + az \qquad ( z\ra 0, \; \; x\ra-\infty) \; ,
$$
\be
\label{47}
\vp(x) \simeq 1 \; - \; \frac{a}{z} \qquad
(z\ra\infty, \; \; x\ra 0) \; .
\ee
It is also convenient to introduce the positively defined function
\be
\label{48}
y \equiv 2 + \vp = y(z) \; ,
\ee
satisfying the boundary conditions
\be
\label{49}
y(0) = 1 \; , \qquad y(1) = 2 \; .
\ee
Then Eq. (41) transforms into
\be
\label{50}
2z^2 \; \frac{d^2y}{dz^2} + 2z \; \frac{dy}{dz} + 6 -
11 y + 6y^2 - y^3 = 0 \; .
\ee

Looking for the form of $y(z)$ at small $z\ra 0$, we substitute
the asymptotic expansion
\be
\label{51}
y(z) \simeq \sum_n a_n \; z^n \qquad (z\ra 0)
\ee
into Eq. (50), from where it follows that
\be
\label{52}
a_0 = 1 \; , \qquad a_n = \frac{(-1)^{n-1}}{2^{n-1}}\;
a_1^n \qquad ( n\geq 1) \; .
\ee
Constructing the factor approximants from expansion (51), we obtain
the same expression
\be
\label{53}
y_k^*(z) = \frac{2 + 3a_1 z}{2 + a_1 z} \qquad ( k\geq 4)
\ee
for orders $k\geq 4$, with the constant $a_1=2$ found from the
boundary conditions (49). Using relations (46) and (48), according
to which
\be
\label{54}
\vp(x) = y(z(x)) - 2 \; ,
\ee
we come to the expression
\be
\label{55}
\vp_k^*(x) = {\rm tanh} \left ( \frac{x}{\sqrt{\ep}} \right )
\qquad (k \geq 4) \; .
\ee
This is the {\it exact} kink solution of the soliton Eq. (41).

Recall that, if we would try to solve Eq. (41) by means of
perturbation theory with respect to large mass, that is, small
$\ep$, we would have to deal with a rather unpleasant and
cumbersome boundary-layer problem [1--3].

\vskip 5mm

{\bf B. Bell soliton equation}

\vskip 3mm

The nonlinear Schr\"odinger equation for a negative mass $-1/\ep$
is
\be
\label{56}
\frac{\ep}{2} \; \frac{d^2\vp}{dx^2} \; - \; \vp + \vp^3 = 0 \; .
\ee
The function $\vp=\vp(x)$ satisfies the boundary conditions
\be
\label{57}
\vp(-\infty) = 0 \; , \qquad \vp(\infty) = 0 \; .
\ee
From these it follows that $\vp(x)$ is a symmetric function,
\be
\label{58}
\vp(-x) = \vp(x) \; ,
\ee
such that
\be
\label{59}
\lim_{x\ra\pm 0} \; \frac{d\vp}{dx} = 0 \; .
\ee
In the vicinity of the stationary points $\vp(\mp\infty)=0$ we have
$$
\vp(x) \simeq a\; \exp \left ( \sqrt{\frac{2}{\ep}}\; x  \right )
\qquad (x\ra -\infty) \; ,
$$
\be
\label{60}
\vp(x) \simeq a\; \exp \left ( -\; \sqrt{\frac{2}{\ep}}\; x \right )
\qquad (x\ra \infty) \; .
\ee
Hence the appropriate convenient variable here is
\be
\label{61}
z \equiv \exp \left ( \sqrt{\frac{2}{\ep}}\; x \right ) \; .
\ee
In terms of the latter, Eq. (60) simplify to
$$
\vp(x) \simeq az \qquad (z\ra 0, \; \; x\ra -\infty) \; ,
$$
\be
\label{62}
\vp(x) \simeq \frac{a}{z} \qquad (z\ra \infty, \; \; x\ra \infty) \; .
\ee

For the function
\be
\label{63}
y = \vp(x(z)) = y(z)
\ee
of variable (61), from Eq. (56), one gets the equation
\be
\label{64}
z^2 \; \frac{d^2 y}{dz^2} + z\; \frac{dy}{dz} \; - \; y +
y^3 = 0 \; ,
\ee
with the boundary conditions
\be
\label{65}
y(0) = 0 \; , \qquad y(\infty) = 0 \; .
\ee
And property (59) takes the form
\be
\label{66}
\lim_{z\ra 1} \; \frac{dy}{dz} = 0 \; .
\ee

In the asymptotic region of small $z$, the solution to Eq. (64) is
\be
\label{67}
y(z) \simeq \sum_n a_n \; z^n \qquad (z\ra 0) \; ,
\ee
in which
$$
a_{2n} = 0 \qquad (n=0,1,2,\ldots ) \; ,
$$
\be
\label{68}
a_{2n+1} = \left ( -\; \frac{1}{8} \right )^n a_1^{2n+1} \; .
\ee
The corresponding factor approximants take the same form
\be
\label{69}
y_k^*(z) = \frac{8a_1 z}{8+(a_1z)^2} \qquad (k \geq 3)
\ee
for orders $k\geq 3$, with the parameter $a_1=2\sqrt{2}$ given by
the boundary condition (66). Resorting to relations (61) and (63),
we obtain
\be
\label{70}
\vp_k^*(x) =\sqrt{2}\; {\rm sech}\left ( \sqrt{\frac{2}{\ep} }\;
x \right ) \qquad (k \geq 3) \; ,
\ee
which is the {\it exact} solution of Eq. (56) describing a bell
soliton.

\section{Evaluation of approximation accuracy}

In the previous sections, we have considered several examples of
initial-value and boundary-value problems, for which the self-similar
factor approximants result in exact solutions. This fact, as such,
that starting with approximations, one can get exact solutions, is
highly nontrivial. This is why we have focused our attention on its
thorough illustration. But, certainly, the more general situation is
when an exact solution cannot be reconstructed, or just does not
exist at all. How then could we evaluate the accuracy of our
approximate solutions?

\vskip 5mm

{\bf A. Solutions defects and errors}

\vskip 3mm

Suppose a differential equation, that can be represented in the
operator form as
$$
E[\;y(x)\;] = 0 \; ,
$$
defines a function $y(x)$ of a variable $x$ in the interval
$[x_1,x_2]$. The latter can be finite or infinite. Let us find an
approximate solution $y_k^*(x)$ of the $k$-th order. In the theory
of differential equations [54,55], one characterizes the accuracy
of approximate solutions in two ways, by calculating the solution
defects and solution errors.

The {\it solution defect} of $y_k^*(x)$ is
\be
\label{71}
D[y_k^*(x)] \equiv |\; E[ y_k^*(x) ]\; | \; .
\ee
This is a local characteristic of an approximate solution $y_k^*(x)$,
showing to what extent the considered solution $y_k^*(x)$ does not
satisfy the given equation $E[y(x)]=0$. For the exact solution $y(x)$,
the solution defect is, evidently, zero.

Varying $x$ in the whole interval $[x_1,x_2]$, one defines the {\it
maximal solution defect}
\be
\label{72}
D[ y_k^* ] \equiv \sup_x D[ y_k^*(x) ] \; ,
\ee
which is the global characteristic of the validity of the
approximate solution $y_k^*(x)$, with respect to the given equation
$E[y(x)]=0$, in the whole interval of the variable $x\in[x_1,x_2]$.

The {\it solution error} of $y_k^*(x)$ is defined as
\be
\label{73}
\Dlt [ y_k^*(x) ] \equiv |y_k^*(x) - y(x) | \; .
\ee
This local characteristic shows how much the approximate solution
$y_k^*(x)$ deviates from the exact solution $y(x)$ at each $x$.

The {\it maximal solution error}
\be
\label{74}
\Dlt [ y_k^* ] \equiv \sup_x \Dlt [ y_k^*(x) ]
\ee
gives the global characteristic of the maximal deviation of
$y_k^*(x)$ from $y(x)$ in the total interval $[x_1,x_2]$.

Both the solution defects and solution errors characterize the
accuracy of approximate solutions. The explicit relation between
these characteristics depends on the explicit form of the given
differential equation. Also [54,55], one can define the {\it 
error-to-defect ratio}
\be
\label{75}
\dlt [y_k^* ] \equiv \frac{\Dlt[y_k^*]}{D[y_k^*]} \; .
\ee

\vskip 5mm

{\bf B. Boundary layer problem}

\vskip 3mm

To illustrate the definitions of the previous subsection in the
case of the factor approximants, let us consider the boundary-layer
problem given by the equation
\be
\label{76}
\ep \; \frac{d^2 y}{dx^2} + x\; \frac{dy}{dx} \; - \; xy = 0 \; ,
\ee
where the function $y=y(x)$ is defined for $x$ in the interval
$0\leq x\leq 1$, with the boundary conditions
\be
\label{77}
y(0) = 0 \; , \qquad y(1) = e \; .
\ee
Equation (76) is an example of a nontrivial boundary-layer problem
[2,3] for which the conventional matching technique contains not
only powers of $\ep$, but also powers of $\sqrt{\ep}$, and $\ln\ep$,
which makes the matching rather complicated. Also, perturbation
theories in terms of $\ep$ do not allow for the exact validity of
the boundary conditions (77), but the latter are satisfied only
approximately, for small $\ep\ll 1$, making such theories [2,3,12]
inapplicable for large $\ep$. Contrary to this, the method of
factor approximants, being very simple, gives approximate solutions
for arbitrary $\ep$, with high accuracy, and satisfies the boundary
conditions exactly for any $\ep$.

It is convenient, first, to redefine the sought function as
\be
\label{78}
z \equiv y e^{-x} = z(x) \; ,
\ee
in order that the latter be varying between zero and one. For function
(78), problem (76) transforms to
\be
\label{79}
\frac{d^2 z}{dx^2} + \left ( 2 + \frac{x}{\ep} \right )
\frac{dz}{dx} + z = 0 \; ,
\ee
with the boundary conditions
\be
\label{80}
z(0) = 0 \; , \qquad z(1) = 1 \; .
\ee
Then we follow the standard procedure, by deriving an expansion of
$z(x)$ for asymptotically small $x\ra 0$, and by constructing the
factor approximants based on the latter expansion. Finally, using
relation (78), we obtain the solutions of Eq. (76) as factor
approximants $y_k^*(x)$.

The solution defects (71) for several factor approximants
$y_k^*(x)$ are shown in Fig. 1, which demonstrates good uniform
convergence of the method. Figure 2 shows the solution errors (73)
for the same factor approximants. To better investigate the relation
between different accuracy characteristics, we present the maximal
solution defects (72), maximal solution error (74), and the
error-to-defect ratio (75) in Tables 1 and 2. For small $\ep$,
the numerical convergence is slower than for large $\ep$. Therefore,
in Table 1 for $\ep=0.1$, we present the accuracy of higher
approximants up to $k=17$. Numerical convergence for $\ep\geq 1$
is so fast that it is sufficient to consider the approximants up
to $k=7$, as in Table 2. As is seen, the maximal solution errors
are much smaller than the maximal solution defects. Hence the
latter can serve as an upper bound for the former.

\section{Nonlinear radial equations}

Now we shall illustrate the power of the method for some physically
motivated nonlinear equations in radial variables.

\vskip 5mm

{\bf A. Gross-Pitaevskii equation}

\vskip 3mm

This is the nonlinear Schr\"odinger equation applied to diferent
Bose-condensed systems (see Refs. [56--62]). We consider here the
variant of this equation describing vortices in Bose systems.
Analogous equations describe also vortices in superfluids,
superconductors, and Higgs fields. The equation, in dimensionless
units, reads as
\be
\label{81}
\frac{d^2\vp}{dr^2} + \frac{1}{r} \; \frac{d\vp}{dr} \; - \;
\frac{\vp}{r^2} + \vp - \vp^3 = 0 \; ,
\ee
defining a function $\vp=\vp(r)$ of the radial variable $r\geq 0$.
The boundary conditions are
\be
\label{82}
\vp(0) = 0 \; , \qquad \vp(\infty) = 1 \; .
\ee

The asymptotic behavior of the solution to Eq. (81) at small $r$ is
\be
\label{83}
\vp(r) \simeq \sum_n a_{2n+1} r^{2n+1} \qquad (r\ra 0) \; ,
\ee
where
$$
a_3 = -\; \frac{1}{8}\; a_1 \; , \qquad
a_5 = \frac{1+8a_1^2}{192}\; a_1 \; ,
$$
$$
a_7 = -\; \frac{1+80a_1^2}{9216}\; a_1\; , \qquad
a_9 = \frac{1+656a_1^2+1152a_1^4}{737280}\; a_1 \; ,
$$
and so on, with $a_1$ to be defined later from the second of the
boundary conditions (82). For convenience, expansion (83) can be
rewritten as
\be
\label{84}
\vp(r) \simeq a_1 r \; \sum_n b_n r^{2n} \qquad
\left ( r^2 \ra 0 \right ) \; ,
\ee
where $b_n\equiv a_{2n+1}/a_1$. Expression (84) shows that
actually, we have an expansion in powers of $r^2$. The corresponding
factor approximants have the form
\be
\label{85}
\vp_k^*(r) = c_k r \prod_i \left ( 1 + A_i r^2
\right )^{n_i} \; .
\ee
The boundary condition
\be
\label{86}
\vp_k^*(0) = 0
\ee
for Eq. (85) does not impose additional constraints, being always valid.
And the boundary condition
\be
\label{87}
\vp_k^*(\infty) = 1
\ee
imposes on Eq. (85) two constraints
\be
\label{88}
c_k \; \prod_i A_i^{n_i} = 1 \; , \qquad
1 + 2 \sum_i n_i = 0 \; .
\ee

In labelling the factor approximants, we associate the order $k$ with
the number of terms in expansion (84) with respect to $r^2$. Then for
the second-order approximant $\vp_2^*$, defined by Eq. (85), we have
$$
c_2=0.518840\; , \qquad A_1=1 \; , \qquad A_2 =0.287401\; ,
$$
$$
n_1=0.026243\; , \qquad n_2=-0.526243 \; .
$$
For the third-order approximant $\vp_3^*(r)$, we get
$$
c_3=0.585667\; , \qquad A_1=0.107803\; , \qquad A_2=0.139245\; ,
$$
$$
n_1=1.761220\; , \qquad n_2=-2.261220 \; .
$$
The fourth-order approximant $\vp_k^*(r)$ is defined by
$$
c_4=0.585331\; , \qquad A_1 = 1\; , \qquad  A_2=0.117323\; ,
$$
$$
A_3=0.130445\; , \qquad n_1=-0.000031\; , \qquad n_2=4.553526\; ,
\qquad n_3=-5.053495 \; .
$$
For the fifth-order approximant $\vp_5^*(r)$, we find
$$
c_5=0.583142 \; , \qquad A_1=0.158576 \; , \qquad
A_2=0.045537+i\; 0.011910 \; , \qquad A_3=A_2^* \; ,
$$
$$
n_1=-0.994835 \; , \qquad n_2=0.247422-i\; 0.429221 \; ,
\qquad n_3=n_2^* \; .
$$
Similarly, any higher-order approximant can be defined following
the standard procedure of Sec. 2. Numerical convergence is quite
fast, which is shown in Table 3. Note that the approximate solutions,
for the same Eq. (81), found  by employing the self-similar root
approximants [27], are less accurate than the self-similar factor
approximants, considered here. To compare the accuracy of the factor
approximants $\vp_k^*(r)$ and the root approximants $R_k^*(r)$,
given in Appendix A, we present in Table 3 both the maximal solution
defects $D[\vp_k^*]$ as well as $D[R_k^*]$. As is seen, the factor
approximants are two orders more accurate than the root approximants.

\vskip 5mm

{\bf B. Stokes-Oseen equation}

\vskip 3mm

A simplified spherically symmetric variant of the Stokes-Oseen
equation [1--3] can be written in the form
\be
\label{89}
\frac{d^2 u}{dr^2} + \frac{2}{r}\; \frac{du}{dr} +
\ep u \; \frac{du}{dr} = 0 \; ,
\ee
where $u=u(r)$ and $r\geq 1$. This equation describes the viscous
flow past a sphere of unit radius, with $\ep$ playing the role of the
Reynolds number. The boundary conditions are
\be
\label{90}
u(1) = 0 \; , \qquad u(\infty) = 1 \; .
\ee
When considering this equation by means of perturbation theory with
respect to the Reynolds number $\ep$, one confronts a very delicate
singular boundary-layer problem [2,3], with complicated expansions
and matching, involving unexpected orders such as $\ep\ln(1/\ep)$.
But in our approach, the problem is easily solvable.

It is convenient, first, to change the variable to
\be
\label{91}
x \equiv r -1 \; .
\ee
The function
\be
\label{92}
y \equiv u(r(x)) = y(x)
\ee
obeys the equation
\be
\label{93}
\frac{d^2 y}{dx^2} + \frac{2}{1+x} \; \frac{dy}{dx} +
\ep y \; \frac{dy}{dx} = 0 \; ,
\ee
with the boundary conditions
\be
\label{94}
y(0) = 0 \; , \qquad y(\infty) = 1 \; .
\ee
Then, as usual, we construct the factor approximants $y_k^*(x)$ and,
using relations (91) and (92), return to the factor approximants
$u_k^*(r)$. The maximal defects of the latter are given in Table 4
for different $\ep$. For small Reynolds numbers $\ep$, the numerical
convergence is very fast. For $\ep\sim 1$, convergence is a little
slower, though again it is easy to reach high accuracy. It is feasible
to reach quite good accuracy even for large $\ep\gg 1$, although this
requires to construct the factor approximants of higher orders
$k\sim 10$.

\vskip 5mm

{\bf C. Strongly-singular problem}

\vskip 3mm

It is instructive to analyze the equation
\be
\label{95}
\frac{d^2 u}{dr^2} + \frac{1}{r}\; \frac{du}{dr} +
\left ( \frac{du}{dr} \right )^2 + \ep u\; \frac{du}{dr} = 0 \; ,
\ee
defining the function $u(r)$ for $r\geq 1$, with the same boundary
conditions
$$
u(1) = 0 \; , \qquad u(\infty) = 1 \; ,
$$
as in Eq. (90). If this equation is treated by perturbation theory
[2,3] with respect to $\ep$, it becomes a terribly complicated
problem. The matching procedure for that equation is notoriously
difficult, because an infinite number of terms, with respect to $\ep$,
must be calculated before even the leading order could be explicitly
matched [2,3].

In the method of the self-similar factor approximants, the procedure
is straightforward, as is described in Sec. 2. We follow the same way
as in the previous subsection. A little difference from the previous
problem (89) is that for Eq. (95) not all even-order approximants exist,
but when they exist, they are close to the odd-order approximants,
similarly to the case of Eq. (89). The odd-order approximants
exist for all orders we have checked and their maximal defects are
presented in Table 5 for different $\ep$, demonstrating good numerical
convergence and the possibility of reaching quite good accuracy.

\section{Conclusion}

We have shown that the method of self-similar factor approximants
[35--37,40,50] can serve as a powerful tool for constructing
approximate solutions to ordinary differential equations. These can
be the initial-value and boundary-value problems, linear as well as
nonlinear differential equations.

A very nontrivial fact is that in some cases factor approximants
automatically reconstruct exact solutions even to nonlinear
equations, provided such solutions exist. This happens, when
the latter solutions can be represented, with the appropriate
change of variables, as factor approximants, that is, when the
solutions pertain to the class of exactly reproduceable functions
[35,50].

The procedure of constructing the factor approximants is very simple,
which is described in Sec. 2. This procedure also is uniquely defined.
If the asymptotic expansion of a solution has the form of Eq. (1),
then all factor approximants enjoy the same structure
$$
f_k^*(x) = f_0(x) \; \prod_i ( 1 + A_i x)^{n_i} \; ,
$$
whose parameters are defined by the accuracy-through-order procedure.

The quantities $A_i$ and $n_i$ can be functions of all other parameters
entering the considered equation. This is why the use of the factor
approximants is not limited to the case of small parameters, but is valid
for any values of the latter, yielding approximate solutions of high
accuracy.

As has been proved in our previous publications [35--37,40,50,63] the
self-similar approximants guarantee an essentially higher accuracy than
Pad\'e approximants [64]. This fact can be easily understood, if one
notices that Pad\'e approximants have the structure which is just a
particular variant of the factor approximants. In addition, the factor
approximants, contrary to Pad\'e approximants, enjoy a great
advantage of being uniquely defined.

Finally, since the quantities $A_i$ and $n_i$ in the expression for
the factor approximants can be functions of other parameters, they
also can be functions of other variables, entering the equations,
and even functions of external fields. This could open the way for
extending the suggested technique to solving partial and stochastic
differential equations [65]. The latter, however, is the subject for
future investigations.

\newpage

{\bf{\Large Appendix A}}

\vskip 5mm

In Sec. 6, subsection 6.1, the factor-approximant solutions
$\vp_k^*(r)$ to the Gross-Pitaevskii equation (81) are compared with
the root approximant solutions $R_k^*(r)$. In Table 3, the maximal
defects of both types of approximants are given.

The self-similar root approximants [26,27,29,30] are constructed from
the asymptotic expansion
$$
y(r) \simeq 1 \; - \; \frac{1}{2}\; r^{-2} \; - \;
\frac{9}{8}\; r^{-4} \; - \; \frac{161}{16}\; r^{-6}
$$ for large $r\ra\infty$, which is an expansion in powers of $r^{-2}$.
The first several root approximants are
$$
R_2^*(r) = \frac{r}{2} \left ( 1 + \frac{1}{4}\; r^2
\right )^{-1/2} \; ,
$$
$$
R^*_3(r) = \frac{r}{\sqrt{2}} \left ( 1 + \frac{1}{2}\; r^2 +
\frac{1}{4}\; r^4 \right )^{-1/4} \; ,
$$
$$
R^*_4(r) = \frac{r}{4^{1/3}} \left ( 1 + \frac{3}{4}\; r^2 +
\frac{3}{16}\; r^4  + \frac{1}{16}\; r^6 \right )^{-1/6} \; ,
$$
$$
R^*_5(r) = \frac{r}{136^{1/8}} \left ( 1 + r^2 +
\frac{9}{68}\; r^4  + \frac{1}{34}\; r^6 + \frac{1}{136}\; r^8
\right )^{-1/8} \; .
$$
The maximal solution defect $D[R_k^*]$ happens at $r\approx 1$ and is
shown in Table 3. Though the accuracy of the root approximants is not
bad, it is two orders lower than that of the factor approximants.

\newpage

\begin{center}

{\large{\bf Figure Captions}}

\end{center}

\vskip 5mm

{\bf Fig. 1}. Solution defects $D[y_k^*(x)]$ of several factor
approximants $y_k^*(x)$ with $\ep=1$, for $k=4$ (solid line), $k=5$
(dashed line), $k=6$ (dotted line), and $k=7$ (dashed-dotted line), for
Eq. (76).

\vskip 5mm

{\bf Fig. 2}. Solution errors $\Dlt[y_k^*(x)]$ of the same factor
approximants, as in Fig. 1, for the same orders $k=4$ (solid line),
$k=5$ (dashed line), $k=6$ (dotted line), and $k=7$ (dashed-dotted line),
for Eq. (76).

\newpage

\begin{center}

{\large{\bf Table Captions}}

\end{center}

{\bf Table 1}. Maximal solution defects $D[y_k^*]$, maximal solution
errors $\Dlt[y_k^*]$, and the error-to-defect ratio $\dlt[y_k^*]$ for
Eq. (76), with $\ep=0.1$.

\vskip 5mm

{\bf Table 2}.  Maximal solution defects $D[y_k^*]$, maximal solution
errors $\Dlt[y_k^*]$, and the error-to-defect ratio $\dlt[y_k^*]$ for
Eq. (76), with $\ep=1$ and $\ep=10$.

\vskip 5mm

{\bf Table 3}.  Maximal solution defects $D[\vp_k^*]$ of the factor
approximants $\vp_k^*(r)$ and maximal solution defects $D[R_k^*]$ of the
root approximants $R_k^*(r)$ for Eq. (81).

\vskip 5mm

{\bf Table 4}. Maximal solution defects $D[u_k^*]$ of the factor
approximants $u_k^*(r)$ for different Reynolds numbers $\ep$, in the case
of Eq. (89).

\vskip 5mm

{\bf Table 5}. Maximal solution defects $D[u_k^*]$ of the factor
approximants $u_k^*(r)$ for different $\ep$, in the case of the strongly
singular Eq. (95).

\newpage

\vskip 5mm

\begin{figure}[h]
\epsfig{figure=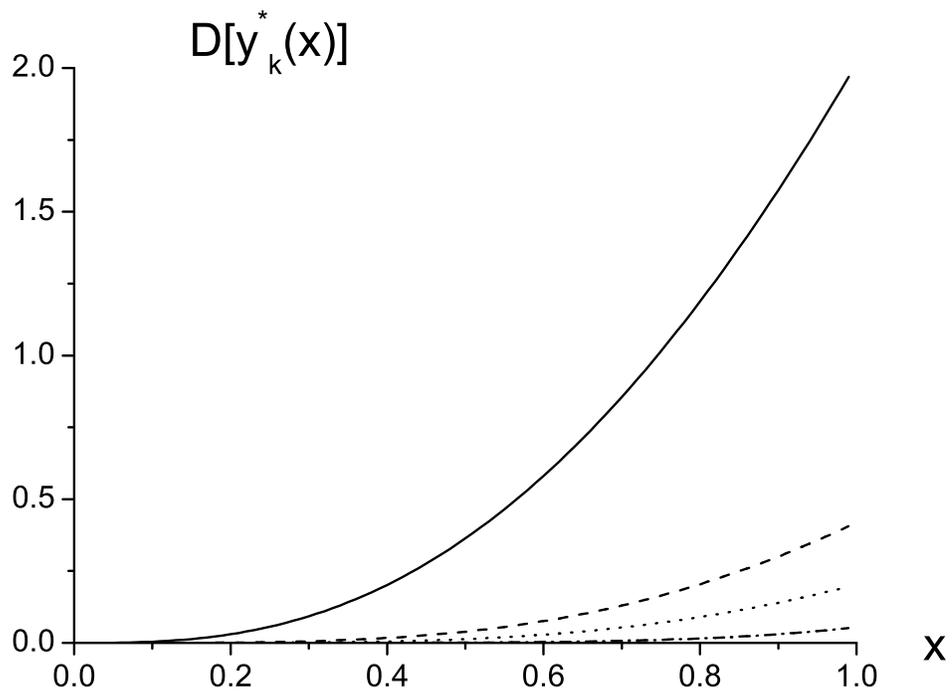,width=14cm}
\caption{Solution defects $D[y_k^*(x)]$ of several factor
approximants $y_k^*(x)$ with $\ep=1$, for $k=4$ (solid line), $k=5$
(dashed line), $k=6$ (dotted line), and $k=7$ (dashed-dotted line), for
Eq. (76).}
\label{Fig.1}
\end{figure}

\newpage

\begin{figure}[h]
\epsfig{figure=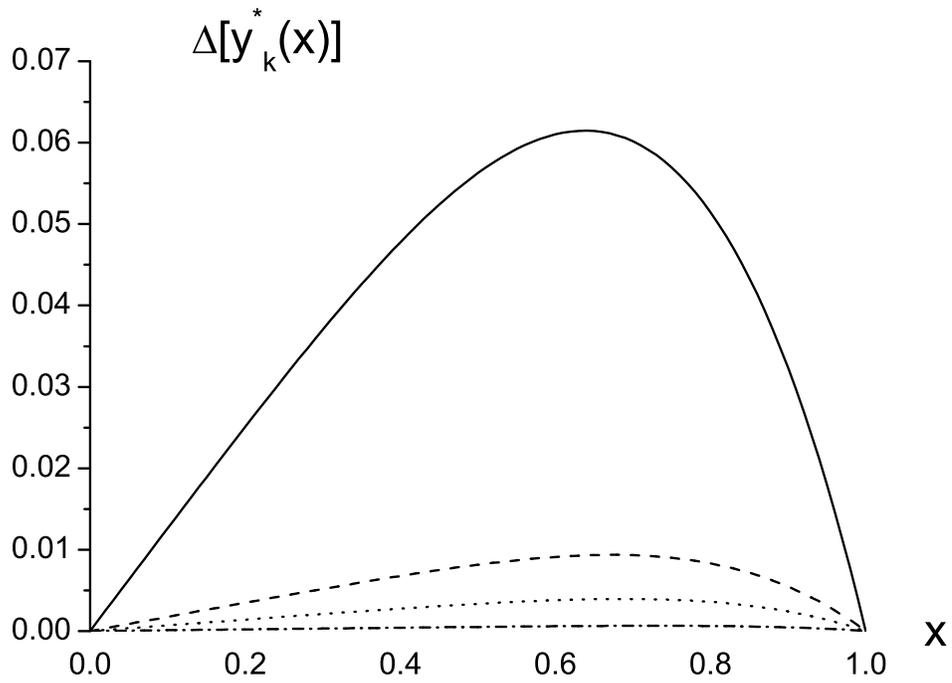,width=14cm}
\caption{Solution errors $\Dlt[y_k^*(x)]$ of the same factor
approximants, as in Fig. 1, for the same orders $k=4$ (solid line),
$k=5$ (dashed line), $k=6$ (dotted line), and $k=7$ (dashed-dotted line),
for Eq. (76).}
\label{Fig.2}
\end{figure}

\newpage

\begin{tabular}{|c|c|c|c|} \hline
$ k $& $ D[y^*_k]$  & $\Dlt[y^*_k]$ & $\dlt[y^*_k]$  \\ \hline
8    &  1.8       &  0.14           & 0.078 \\ \hline
9    &  0.58      &  0.051          & 0.088 \\ \hline
10   &  0.20      &  0.018          & 0.090 \\ \hline
11   &  0.68      &  0.038          & 0.056 \\ \hline
12   &  0.28      &  0.015          & 0.054 \\ \hline
13   &  0.090     &  0.0048         & 0.053 \\ \hline
14   &  0.024     &  0.0014         & 0.058 \\ \hline
15   &  0.057     &  0.0022         & 0.039 \\ \hline
16   &  0.032     &  0.0012         & 0.038 \\ \hline
17   &  0.0083    &  0.00028        & 0.034 \\ \hline
\end{tabular}

\vskip 0.5cm
{\parindent=0pt
{\bf Table 1}: Maximal solution defects $D[y_k^*]$, maximal solution
errors $\Dlt[y_k^*]$, and the error-to-defect ratio $\dlt[y_k^*]$ for
Eq. (76), with $\ep=0.1$.}

\vskip 3cm

\begin{tabular}{|c|c|c|c|c|c|c|} \hline
    &   & $\ep=1$ &    &    & $\ep=10$ &     \\ \hline
$k$ & $D[y_k^*]$ & $\Dlt[y_k^*]$ & $\dlt[y_k^*]$ & $D[y_k^*]$ &
$\Dlt[y_k^*]$ & $\dlt[y_k^*]$ \\ \hline
4   & 2.0  & 0.062  & 0.031  &  1.58   &  0.0051   & 0.0032  \\ \hline
5   & 0.42 & 0.0094 & 0.022  &  0.059  &  0.00013  & 0.0022   \\ \hline
6   & 0.20 & 0.0040 & 0.020  &  0.025  &  0.000046 & 0.018 \\ \hline
7   & 0.055& 0.00064& 0.012  &  0.0032 & 3.8$\cdot 10^{-6}$ & 0.012 \\
\hline
\end{tabular}

\vskip 0.5cm

{\parindent=0pt
{\bf Table 2}: Maximal solution defects $D[y_k^*]$, maximal solution
errors $\Dlt[y_k^*]$, and the error-to-defect ratio $\dlt[y_k^*]$ for
Eq. (76), with $\ep=1$ and $\ep=10$.}

\newpage

\begin{tabular}{|c|c|c|} \hline
$ k$ & $D[y_k^*]$ & $D[R_k^*]$ \\ \hline
2    & 0.12       & 0.14  \\ \hline
3    & 0.017      & 0.25 \\ \hline
4    & 0.015      & 0.10 \\ \hline
5    & 0.0020     & 0.11 \\ \hline
6    & 0.0018     & 0.10 \\ \hline
\end{tabular}

\vskip 0.5cm

{\parindent=0pt
{\bf Table 3}: Maximal solution defects $D[\vp_k^*]$ of the factor
approximants $\vp_k^*(r)$ and maximal solution defects $D[R_k^*]$ of the
root approximants $R_k^*(r)$ for Eq. (81).}

\vskip 3cm

\begin{tabular}{|c|c|c|c|} \hline
$ k $& $ \ep=0.1  $ & $ \ep=1  $    & $ \ep=10 $ \\ \hline
4    &  0.86     &  0.094     & 3.1        \\ \hline
5    &  0.0035   &  0.015     & 12    \\ \hline
6    &  0.00068  &  0.024     & 40    \\ \hline
7    &  0.00015  &  0.023     & 12    \\ \hline
8    &  0.00010  &  0.00095   & 0.32  \\ \hline
9    &  0.000056 &  0.00095   & 0.15    \\ \hline
10   &  0.000056 &  0.0015    & 0.15   \\ \hline
11   &  0.000026 &  0.0015    & 0.23   \\ \hline
12   &  0.000024 &  0.0015    & 0.011 \\ \hline
\end{tabular}

\vskip 0.5cm
{\parindent=0pt
{\bf Table 4}: Maximal solution defects $D[u_k^*]$ of the factor
approximants $u_k^*(r)$ for different Reynolds numbers $\ep$, in the case
of Eq. (89)}

\vskip 3cm

\begin{tabular}{|c|c|c|c|} \hline
$ k $& $ \ep=0.1  $ & $ \ep=1  $    & $ \ep=10  $ \\ \hline
3    &  0.0035      &  0.062        & 1.4        \\ \hline
5    &  0.00056     &  0.080        & 0.44     \\ \hline
7    &  0.00060     &  0.011        & 0.54        \\ \hline
9    &  0.00035     &  0.0017       & 0.32       \\ \hline
11   &  0.00015     &  0.0017       & 0.078     \\ \hline
\end{tabular}

\vskip 0.5cm
{\parindent=0pt
{\bf Table 5}: Maximal solution defects $D[u_k^*]$ of the factor
approximants $u_k^*(r)$ for different $\ep$, in the case of the
strongly singular Eq. (95).}


\newpage

\begin{thebibliography}{99}
\bibitem{1}
G.E.O. Giacaglia, Perturbation Methods in Nonlinear Systems,
Springer, New York, 1972.

\bibitem{2}
A.H. Nayfeh, Perturbation Methods, Wiley, New York, 1973.

\bibitem{3}
E.J. Hinch, Perturbation Methods, Cambridge University,
Cambridge, 1991.

\bibitem{4}
H. Haken, Advanced Synergetics, Springer, Berlin, 1982.

\bibitem{5}
I.K. Kudryavtsev, Chemical Instabilities, Moscow University,
Moscow, 1987.

\bibitem{6}
W. Weidlich, Phys. Rep. 204 (1991) 1.

\bibitem{7}
D. Sornette, Critical Phenomena in Natural Sciences,
Springer, Berlin, 2006.

\bibitem{8}
V.I. Yukalov, Moscow Univ. Phys. Bull. 31 (1976) 10.

\bibitem{9}
V.I. Yukalov, E.P. Yukalova, Ann. Phys. (N.Y.) 277
(1999) 219.

\bibitem{10}
F. Cooper, H.K. Shepard, L.M. Simmons, Phys. Lett. A 156
(1991) 436.

\bibitem{11}
F. Cooper, L.M. Simmons, P. Sodano, Physica D 56 (1992) 68.

\bibitem{12}
L.Y. Chen, N. Goldenfeld, Y. Oono, Phys. Rev. E 54 (1996)
376.

\bibitem{13}
T. Hatsuda, T. Kunihiro, T. Tanaka, Phys. Rev. Lett. 78
 (1997) 3229.

\bibitem{14}
T. Kunihiro, Phys. Rev. D 57 (1998) 2035.

\bibitem{15}
S.I. Ei, K. Fujii, T. Kunihiro, Ann. Phys. (N.Y.) 280
(2000) 236.

\bibitem{16}
A. Okopinska, Phys. Rev. E 65 (2002) 062101.

\bibitem{17}
P. Amore, A. Aranda, Phys. Lett. A 316 (2003) 218.

\bibitem{18}
V.I. Yukalov, Phys. Rev. A 42 (1990) 3324.

\bibitem{19}
V.I. Yukalov, Physica A 167 (1990) 833.

\bibitem{20}
V.I. Yukalov, J. Math. Phys. 32(1991) 1235.

\bibitem{21}
V.I. Yukalov, J. Math. Phys. 33 (1992) 3994.

\bibitem{22}
V.I. Yukalov, E.P. Yukalova, Physica A 225 (1996) 336.

\bibitem{23}
V.I. Yukalov, S. Gluzman, Phys. Rev. Lett. 79 (1997) 333.

\bibitem{24}
S. Gluzman,  V.I. Yukalov, Phys. Rev. E 55 (1997) 3983.

\bibitem{25}
V.I. Yukalov, S. Gluzman, Phys. Rev. E 55 (1997) 6552.

\bibitem{26}
V.I. Yukalov, E.P. Yukalova, Chaos Solit. Fract. 14 (2002) 839.

\bibitem{27}
V.I. Yukalov, E.P. Yukalova, S. Gluzman, Phys. Rev. A 58
 (1998) 96.

\bibitem{28}
V.I. Yukalov, S. Gluzman, Phys. Rev. E 58 (1998) 1359.

\bibitem{29}
S. Gluzman, V.I. Yukalov, Phys. Rev. E 58 (1998) 4197.

\bibitem{30}
V.I. Yukalov, S. Gluzman, Physica A 273 (1999) 401.

\bibitem{31}
S. Gluzman, D. Sornette, Phys. Rev. E 63 (2001) 066129.

\bibitem{32}
S. Gluzman, J.V. Andersen, D. Sornette, Comput. Seismology 32
 (2001) 122.

\bibitem{33}
V.I. Yukalov, E.P. Yukalova, V.S. Bagnato, Phys. Rev. E 66
 (2002) 025602.

\bibitem{34}
D. Sornette, A. Helmstetter, J.V. Andersen, S. Gluzman, J.R. Grasso,
V. Pisarenko, Physica A 338 (2004) 605.

\bibitem{35}
S. Gluzman, V.I. Yukalov,  and D. Sornette, Phys. Rev. E 67
(2003) 026109.

\bibitem{36}
V.I. Yukalov, S. Gluzman, D. Sornette, Physica A 328
(2003) 409.

\bibitem{37}
V.I. Yukalov, S. Gluzman, Int. J. Mod. Phys. B 18 (2004) 3027.

\bibitem{38}
D. Park, Physica 22 (1956) 932.

\bibitem{39}
C.J. Thompson, A.J. Guttmann, B.W. Ninham, J. Phys. C 2
(1969) 1889.

\bibitem{40}
V.I. Yukalov, E.P. Yukalova, Eur. Phys. J. B 55 (2007) 93.

\bibitem{41}
R.R.P. Singh, S. Chakravarty, Phys. Rev. B 36 (1987) 546.

\bibitem{42}
R.R.P. Singh, S. Chakravarty, Phys. Rev. B 36 (1987) 559.

\bibitem{43}
D.P. Landau, J. Magn. Magn. Mater. 200 (1999) 231.

\bibitem{44}
M. Campostrini, A. Pelissetto, P. Rossi, E. Vicari, Phys. Rev. E
65 (2002) 066127.

\bibitem{45}
A. Pelissetto, E. Vicari, Phys. Rep. 368 (2002) 549.

\bibitem{46}
M. Campostrini, M. Hasenbusch, A. Pelissetto, E. Vicari, arXiv:
cond-mat/0605083 (2006).

\bibitem{47}
J. Zinn-Justin, Quantum Field Theory and Critical Phenomena,
Oxford University, Oxford, 1996.

\bibitem{48}
H. Kleinert, V. Schulde-Frohlinde, Critical Properties of
$\vp^4$-Theories, World Scientific, New Jersey, 2001.

\bibitem{49}
H. Kleinert, Path Integrals, World Scientific, Singapore, 2006.

\bibitem{50}
V.I. Yukalov, E.P. Yukalova, Phys. Lett. A 368 (2007) 341.

\bibitem{51}
N. Bourbaki, Fonctions d'une Variable R\'eelle, Hermann, Paris,
1965.

\bibitem{52}
U.D. Jentschura, J. Zinn-Justin, Phys. Lett. B 596 (2004) 138.

\bibitem{53}
J.S. Nicolis, Dynamics of Hierarchical Systems, Springer, Berlin,
1986.

\bibitem{54}
H.J. Stetter, Analysis of Discretization Methods for Ordinary
Differential Equations, Springer, Berlin, 1973.

\bibitem{55}
E. Hairer, S.P. Norsett, G. Wanner, Solving Ordinary Differential
Equations, Springer, Berlin, 1987.

\bibitem{56}
L. Pitaevskii, S. Stringari, Bose-Einstein Condensation in Dilute
Gases, Clarendon, Oxford, 2003.

\bibitem{57}
P.W. Courteille, V.S. Bagnato, V.I. Yukalov, Laser Phys. 11
(2001) 659.

\bibitem{58}
J.O. Andersen, Rev. Mod. Phys. 76(2004) 599.

\bibitem{59}
V.I. Yukalov, Laser Phys. Lett. 1 (2004) 435.

\bibitem{60}
K. Bongs, K. Sengstock, Rep. Prog. Phys. 67 (2004) 907.

\bibitem{61}
V.I. Yukalov, M.D. Girardeau, Laser Phys. Lett. 2 (2005) 375.

\bibitem{62}
A. Posazhennikova, Rev. Mod. Phys. 78 (2006) 1111.

\bibitem{63}
S. Gluzman, V.I. Yukalov, J. Math. Chem. 39 (2005) 47.

\bibitem{64}
G.A. Baker, P. Graves-Moris, P\'ade Approximants, Cambridge
University, Cambridge, 1996.

\bibitem{65}
L. Arnold, Random Dynamical Systems, Springer, Berlin, 1998.

\end{thebibliography}
\end{document}